\documentstyle[aps,prd]{revtex}
\begin{document}

\setcounter{page}{1}
\title{Relaxation versus collision times in the cosmological radiative era}
\author{ Diego
Pav\'on$^\ddagger$ and Roberto A. Sussman$^\dagger$}
\address
{$^\ddagger$ Departamento de F\'{\i}sica, Facultad de Ciencias,
Universidad Aut\'onoma de Barcelona, \\
E-08193 Bellaterra, Spain.\\$^\dagger$ Instituto de Ciencias Nucleares,
Apartado Postal 70543, UNAM, M\'exico DF, 04510, M\'exico.\\}

\maketitle
\begin{abstract}
We consider the Lema\^{\i}tre-Tolman-Bondi metric with an inhomogeneous
viscous fluid source satisfying the equation of state of an interactive
mixture of radiation and matter. Assuming conditions prior to the decoupling
era, we  apply Extended Irreversible Thermodynamcs (EIT) to this mixture.
Using the full transport equation of EIT we show that the relaxation time of
shear viscosity can be several orders of magnitude larger than the Thomson
collision time between photons and  electrons. A comparison with the
``truncated'' transport equation for these models reveals that the latter
cannot describe properly the decoupling of matter and radiation.
\end{abstract}

\section {Introduction}

The relaxation time $\tau$ for the decay of any
dissipative flux in a hydrodynamic system is often misunderstood and
uncritically assumed to vanish (an unphysical assumption) or be of
the order of the relevant collision time $t_{col}$ of the particles making
up the fluid -for a recent review see \cite{MDV}. The root of this confusion
may be traced to the great success of the Fourier-Navier-Stokes theory
of dissipative transport \cite{FW}. This venerable theory is very
useful for engineering applications, but it  ignores altogether
the relaxation times taken by the system to return to thermodynamic
equilibrium state once the gradients of temperature, fluid
velocity and so on are suddenly turned off, and therefore
its range of validity is limited (at most) to steady-state
situations \cite{JCL}. Therefore, this theory displays a serious drawback:
the prediction of instantaneous propagation of the dissipative signals which,
on physical grounds, should be limited by the  speed of sound.
This fact becomes important when dealing with
relativistic self--gravitating fluid distributions such as a collapsing
star, as well as for the early Universe (see \cite{Herrera}, \cite{Romano}
and references therein). Obviously $\tau$  and $t_{col}$ ought to be related
to one another, however there is no generic rule on how to find such a
relationship since it depends in each case on the particular system under
consideration.

In a previous paper we studied the matter-radiation decoupling for an
expanding inhomogeneous universe described by new exact solutions
\cite{mixt} characterized by the Lema\^{\i}tre--Tolman--Bondi (LTB) metric with
flat space sections \cite{Bondi,Krasinski} and a viscous fluid source. The
dissipative quantities involved (the shear stresses) in these models are well
behaved and comply with relativistic causality and stability in  the sense
of Israel and Stewart \cite{IS,DDC,HL}. As stressed in \cite{Krasinski},
the study of inhomogeneous cosmologies is a well motivated and justified
endeavor, especially in view of the results of Mustapha {\it et al.} who show
that inhomogeneous but spherically symmetric models can be compatible with
current CBR observations \cite{mustapha}. This paper aims at showing
that the solution found in \cite{mixt} implies that the relaxation
time of the dissipative shear stress of the cosmic matter--radiation 
mixture prior decoupling is several orders of magnitud larger than
the collision time between photons and electrons. In section II we present
our model, recall the exact solution derived in \cite{mixt}, and compare
the aforesaid relaxation time with the collision time between photons and 
electrons before the matter-radiation decoupling is achieved. Section III 
summarizes our conclusions.

\section{The LTB dissipative model}
Cosmic matter sources in the period from the end of cosmic nucleosynthesis to
the decoupling of matter and radiation (roughly the temperature range  $
10^{3} \mbox{K} \leq T \leq 10^{6}\mbox{K}$) can be characterized as a
mixture of ultra-relativistic and non-relativistic particles (``radiation''
and ``matter''), subjected to a tight interaction via various
radiative processes (hence the name ``radiative'' era). The usual approach
for studying this era is to consider a FLRW space-time background where
matter sources are described  by: (a) equilibrium kinetic theory, (b) gauge 
invariant perturbations, or (c) by hydrodynamical models, which, in general, 
fail to incorporate a physically plausible description of the matter-radiation
interaction. Examination of these  matter sources in an inhomogeneous 
hydrodynamic context requires an imperfect fluid whose state
variables are compatible with the equation of state of a  mixture of a
non-relativistic gas (matter) and a extreme relativistic one (radiation).
In \cite{mixt} we derived a class of exact solutions along these lines, so
that: (a) matter-radiation radiative interaction was modeled by a
dissipative shear-stress tensor, (b) only the evolution along the radiative
era was considered, (c) the background space-time was described by the LTB
metric with flat spacelike slices:

\begin{equation}
ds^2= -c^2dt^2+ Y'^{2} \,dr^2 +Y^2\left[d\theta^2+\sin^2
(\theta) d\phi^2
\right]\, , \quad Y = Y(t,r).
\label{ltb}
\end{equation}
\noindent
For a comoving 4-velocity $u^a=c\delta^a_t $ in (\ref{ltb}), the expansion
scalar and shear tensor are
\begin{equation}
u^a_{;a}\equiv\Theta= {\dot Y'\over{Y'}}+{{2\dot Y}\over Y} \, , \qquad
\sigma^a\,_b= {\bf{\hbox{diag}}}\left[0,-2\sigma,\sigma,\sigma
\right],\,\quad \sigma\equiv{1\over 3}\left({\dot Y\over Y}-{\dot Y'\over Y'}
\right)
\label{expansion}
\end{equation}
\noindent
where a prime denotes derivative with respect to $r$ and
$\dot{Y}=u^aY_{,a}=Y_{,t}$.

The stress-energy tensor of the matter--radiation mixture is
\begin{equation}
T^{ab}= \rho u^au^b+ph^{ab}+\Pi^{ab} \qquad (h^{ab}= c^{-2}u^a
u^b+g^{ab},\quad \Pi^a\,_a= 0) \, ,
\label{setensor}
\end{equation}
where the  most general form for the shear-stress for (1) is given by:
${\bf{\hbox{diag}}}\left[0,-2P,P,P \right]$, with $P = P(t,r)
$ to be determined by the field equations. The relation between $\rho$ and $p$
follows by assuming that the matter component is ``dust" (the low temperature
limit of a classical ideal gas, hence its pressure can be neglected) and
therefore
the hydrostatic pressure is furnished by the radiation only, i.e.
\[
\rho\approx mc^2n^{(m)}+3n^{(r)}k_{_B}T, \qquad
p\approx p^{(r)} = n^{(r)}k_{_B}T
\]
\noindent
where the superscripts $(m)$ and $(r)$ denote the particle number densities of
``matter'' and ``radiation''. The shear dissipative pressure tensor is governed
by the transport equation of EIT
\begin{equation}
\tau \dot\Pi_{cd}\,h^c_ah^d_b+\Pi_{ab}\left[
1+{1\over2}T\eta\left({{\tau}\over{T\eta}}  \,
u^c\right)_{;c}\right]+ 2\eta\,\sigma_{ab}= 0, \quad \mbox {with}
\quad
\eta=\eta_{_{(rg)}}= {4\over{5}}p^{(r)}\tau \, ,
\label{pitensor}
\end{equation}
\noindent
where $\tau$ is the relaxation time for the shear--stress and $\eta$ is the
coefficient of shear viscosity. The subscript $_{rg}$ in $\eta$ indicates
that we
use the form of this coefficient provided by kinetic theory for the
electron-photon interaction (the ``radiative gas'' model). This transport
equation is compatible with relativistic causality and
stability, and is supported by kinetic theory, statistical fluctuation theory,
and information theory \cite{JCL,IS,DDC,HL}. In its turn the
expression for $\eta$ is well--known in the literature and it can be derived
from different standpoints \cite{JCL,ddj}. The entropy per particle takes
the form
\[
s= s^{(e)}+\frac{\alpha}{nT}\,\Pi_{ab}\Pi^{ab},\quad \Rightarrow \quad
(snu^a)_{;a}=\dot s nu^a\geq 0,
\]
\noindent
where $\alpha=-\tau/(2\eta_{_{(rg)}})=-5/(8p^{(r)})$,\, $s^{(e)}$ is the
equilibrium photon entropy per baryon, and we have used the fact
that the number of particles of each fluid entering the mixture is
independently conserved: 
$ (n^{(m)}u^a)_{;a} = (n^{(r)}u^a)_{;a} = 0$, as well as
$n^{(m)}/n^{(r)} \ll 1$. The positive--definite nature of $\dot s$  and $\tau$
above is crucial for the theory to comply with relativistic causality,
notwithstanding the unjustified and widespread belief that $\tau$ vanishes
altogether or is at most  of the order of the collision time.

\subsection{Exact Solution and Density Contrasts}
Integration of Einstein's field equations (see details in \cite{mixt})
leads to the following exact solution
\begin{equation}
\frac{3}{2}\sqrt{\mu} \,(t-t_{i})=
\sqrt{y+\epsilon}\left(y-2\epsilon
\right)-
\sqrt{1+\epsilon}\left(1-2\epsilon\right) ,
\label{solution}
\end{equation}
\noindent
where the definitions
\[
\mu\equiv  \frac{\kappa M}{Y_i^3},\qquad \epsilon\equiv
\frac{W}{M},\qquad
y\equiv\frac{Y}{Y_i} \, ,
\]
\noindent
were used along with
\[
M= \int{\rho_i^{(m)}Y_i^2Y'_idr},\qquad \rho_i^{(m)}\equiv mc^2
n_i^{^{(m)}} \, , \qquad
W= \int{\rho_i^{(r)}Y_i^2Y'_idr},\qquad \rho_i^{(r)}\equiv
3n_i^{^{(r)}}k_{_B}T_{i} \, ,
\]
\noindent
so that $\rho_i^{(m)},\,\rho_i^{(r)} $ define the initial
densities of the non-relativistic and  relativistic
components of the mixture, respectively. The subscript ``i"
refers to some initial hypersurface to be specified below, and $\kappa$
denotes Einstein's gravitational constant.

For the inhomogeneous solutions derived above it is convenient to
characterize initial conditions by introducing suitably averaged initial
densities: $\left\langle
\rho_i^{(m)}\right\rangle,\,\left\langle
\rho_i^{(r)}\right\rangle $ and density contrast parameters
$\Delta_i^{^{(m)}},\,\Delta_i^{^{(r)}} $ defined by
\[
3\left\langle
\rho_i^{(m)}\right\rangle Y_i^3=\int{\rho_i^{(m)}\,d(Y_i^3)}, \qquad
3\left\langle
\rho_i^{(r)}\right\rangle Y_i^3=\int{\rho_i^{(r)}\,d(Y_i^3)},
\]
\[
\rho_i^{(m)}=\left\langle
\rho_i^{(m)}\right\rangle\left[1+\Delta_i^{^{(m)}}\right], \qquad
\rho_i^{(r)}=\left\langle
\rho_i^{(r)}\right\rangle\left[1+\Delta_i^{^{(r)}}\right],
\]
\noindent
as well as the  ancillary functions
\[
\Gamma\equiv \frac{Y'/Y}{Y'_i/Y_i}\qquad \Psi\equiv 1+\frac{\left(1-\Gamma
\right)}{3(1+\Delta_i^{^{(r)}})} ,\qquad
\Phi\equiv
1+\frac{\left(1-4\Gamma
\right)}{3(1+\Delta_i^{^{(r)}})} ,
\]
that appear in the functional forms of the state variables:
$n^{(m)}=n_i^{(m)}/(y^3\Gamma) $,\, $n^{(r)}=n_i^{(r)}/(y^3\Gamma) $,\,
$T=T_i\Psi/y$ and $P=(1/2)n_i^{(r)}k_{_B}T_i\Phi/(y^4\Gamma) $. These functions
are linked to the density contrasts by
\begin{equation}
\Gamma= 1+3A\Delta_i^{^{(m)}}+3B\Delta_i^{^{(r)}} \, ,
\label{Gamma}
\end{equation}
\begin{equation}
\Psi= 1-\frac{A\Delta_i^{^{(m)}}
+B\Delta_i^{^{(r)}}}{1+\Delta_i^{^{(r)}}} \, ,
\label{Psi}
\end{equation}
\begin{equation}
\Phi= \frac{-4A\Delta_i^{^{(m)}}
+\left(1-4B\right)\Delta_i^{^{(r)}}}{1+ \Delta_i^{^{(r)}}} \, ,
\label{Phi}
\end{equation}
where the quantities $A$ and $B$ are known functions of $y$
\[
A=\frac{1}{3y^2}\left[\frac{\sqrt{y+\epsilon}}{\sqrt{1+\epsilon}}
(1-4\epsilon-8\epsilon^2)-(y^2-4\epsilon
y-8\epsilon^2)\right] \, ,
\qquad
B=\frac{\epsilon}{y}\left[(1+2\epsilon)\frac{\sqrt{y+\epsilon}}
{\sqrt{1+\epsilon}}-(y+2\epsilon)\right].
\]
\noindent
It should be noted that the solutions must satisfy the restriction
$\Gamma > 0$ so that the particle number densities do not become
negative and that no shell-crossing singularities occur.

At this point it is expedient to point out first that no expression
exists that gives $\tau$ solely in terms of the macroscopic variables
($\rho$, $p$, $T$, n, etc.), i.e., there is not such  thing as an
``equation of state" for $\tau$. Secondly, using the kinetic 
theory of gases (in general) the collision time (not $\tau$) could in
principle be  written as a collision integral, but that expression would
be mathematically  untractable. This is why we resort to (\ref{pitensor})
to get $\tau$ in  terms of the quantities introduced above
\begin{equation}
\tau =\frac{-\Psi\Phi}{\sigma}\; \frac{{\textstyle{{9} \over
{4}}}\left[1+\Delta_i^{^{(r)}}\right]^2}{{\textstyle{{4} \over
{5}}}\left[3+4\Delta_i^{(r)}+{\textstyle{{13} \over
{32}}}\Gamma\right]^2+{\textstyle{{171} \over
{256}}}\Gamma^2 }.
\label{tau}
\end{equation}

\noindent
This expression is justified as long as it behaves as a relaxation parameter
for the interactive matter-radiation mixture in the theoretical framework of
EIT.

Given a set of initial conditions specified by
$\epsilon,\,\Delta_{i}^{^{(m)}},\,\Delta_{i}^{^{(r)}}$,
on some initial hypersurface, the forms of the state and geometric
variables as functions of $y$ and the chosen initial conditions are
fully determined. However, for the solutions to be physically
meaningful they must comply with the following set of physical
restrictions,
\begin{equation}
\dot s= \frac{15 k_{_B}}{16\tau}
\left(\frac{\Phi} {\Psi}\right)^2 > 0,
\label{dots}
\end{equation}
\noindent
(consistent with the condition that the entropy production be
positive),
\begin{equation}
\Psi > 0, \, \quad \sigma \Phi < 0,
\label{psiphi}
\end{equation}
\noindent
to ensure that $\tau$, $p$ and $T$ are all positive. Likewise the
concavity and stability of $s$ demand that
\begin{equation}
\dot \tau>0,\qquad {\ddot s\over\dot
s}= {{2\sigma\Gamma}\over{3\Psi\Phi}}\,{{\left\langle
\rho_i^{(r)}\right\rangle}\over
{\rho_i^{(r)}}}\left[1+{{\left\langle \rho_i^{(r)}\right\rangle}\over
{3\rho_i^{(r)}}}\right]-{\dot\tau\over{\tau}} < 0.
\label{dottau}
\end{equation}

The collision time ($t_\gamma $) for the Thomson scattering of photons by
electrons (the dominant process of shear transport in the radiative  era)
is  given from the Saha equation by

\begin{equation}
t_\gamma ={1 \over {2c\,\sigma _{_T}n^{(m)}}}\left[ {1+\left(
{1+{{4h^3n^{(m)}\exp \left( {B_0/k_{_B}T} \right)} \over {\left( {2\pi
\,m_ek_BT} \right)^{3/2}}}} \right)^{1/2}} \right],
\label{tcol}
\end{equation}
\noindent
where $\sigma _{_T}$, $B_0$, $m_e$ and $h$ are, respectively, the Thomson 
scattering cross section, the hydrogen atom binding energy, the
electron mass and Planck's constant. For details of the derivation of (13)
see \cite{mixt}. Before matter--radiation decoupling $t_{\gamma} < t_{H}$
(where $t_{H} \equiv 3/\Theta$ denotes the expansion time, i.e., the
``Hubble time''). For $t_{\gamma} > t_{H}$ matter and radiation no
longer interact between each other.

Choosing as initial time $t=t_i$ roughly corresponding to $T=10^{6}$ Kelvin, 
the evolution of the models crucially depends on initial conditions 
specified by the useful quantity
$\Delta_{i}^{(s)} \equiv \textstyle{\frac{3}{4}} \Delta_{i}^{(r)} -
\Delta_{i}^{(m)}$ (formally analogous to the definition of adiabatic
perturbations). As shown in \cite{mixt} (see also Figure 3 of this
paper), the collision time $t_{\gamma}$ is quite insensitive to the
values of $\Delta_{i}^{(s)}$ given by initial conditions, therefore for 
some some $y = y_{D}$ we find that $t_{\gamma}$  overcomes $t_{H}$, i.e.
there is always a decoupling hypersurface. Taking the  standard
decoupling temperature $T_{D} \approx 4 \times 10^{3}$, numerical 
evaluation of $T_{D} = (10^{6}/y_{D}) \, \Psi(y_{D}, \Delta_{i}^{(s)},
\Delta_{i}^{(r)})$ where $\Psi$ is given by (\ref{Psi}) leads to $y_{D}
\approx 10^{2.4}$. Also, observational constraints on the decoupling
hypersurface $[\delta T_{eq}/T_{eq}]_{D} \approx 10^{-5}$
further restrict the maximum values of both $\Delta_{i}^{(m)}$ and
$\Delta_{i}^{(r)}$ to about $10^{-4}$. However, $\tau$ is very
sensitive to $\Delta_{i}^{(s)}$, for example, only if $\Delta_{i}^{(s)}=0$
(i.e. for adiabatic initial conditions) $\tau$ has an adequate
asymptotic evolution, in the sense that as $y\to\infty$ (and so,
$t\to\infty$) it grows monotonically (as $\tau\to y^{3/2}$) and 
$\dot{s} \rightarrow 0$ in this limit, though $\tau$ in this case
remains smaller than $t_{H}$ for all the evolution as $y\to\infty$, a
very different behavior from that of $t_{\gamma}$ which overtakes
$t_{H}$ at the decoupling surface. The relaxation time for
adiabatic initial conditions is then unphysical near the decoupling
surface.  For non-adiabatic intial conditions, $\Delta_{i}^{(s)}\neq 0$,
if $\Delta_{(i)}^{(r)} \Delta_{i}^{(s)} > 0$ and $\sigma \Phi < 0$ hold,
we can have $\sigma \rightarrow 0$ for non--vanishing $\Phi$,  therefore
$\tau$ diverges and overtakes $t_{H}$, a behavior that is qualitatively
analogous to that of $t_{\gamma}$. In order to have $\tau$ overtaking
$t_{H}$ at $y_{D}\approx 10^{2.4}$, we find that $\Delta_{i}^{(s)}$
cannot be greater than approximately $10^{-8}$ (quasi-adiabatic initial
conditions). Therefore, under the assumptions of EIT and using the
full transport equation (\ref{pitensor}) and $\tau$ given by (\ref{tau}),
we find that initial conditions do exist so that $\tau$ has a physicaly
reasonable form near the decoupling surface. Figures 2, 3 and 5
illustrate this conclussion.

\subsection{Comparing the different times}
On the initial hypersurfce $\Gamma_{i} = \Psi_{i} =1$,
$\Phi_{i} \approx \Delta_{i}^{(r)}$ and $\tau_{i}$ is evaluated
to be $\tau_{i} \approx 10^{9}$ seconds. Likewise
$[t_{\gamma}]_{i} \approx 2227$ seconds. As a consequence
$[\tau/t_{H}]_{i} \approx 5/16$ while
$[t_{\gamma}/t_{H}]_{i} \approx 10^{-6}$. As the decoupling
hypersurface is approached (i.e., $y \rightarrow y_{D}$)
$\Psi$ and $\Gamma$ slightly depart from unity as depicted in
Figure 1. On that hypersurface one has
$t_{\gamma}(y_{D}) \approx \tau(y_{D}) \sim 10^{13}$ seconds.
Figure 2 shows how sensitive is the latter quantity
to $\Delta_{i}^{(s)}$. We see therefore that contrary
to the usual belief, the model examined in \cite{mixt} shows that
$\tau$ is much larger than  $t_{\gamma}$ for most of the radiative era
and the timescales $\tau,\,t_\gamma$  become comparable only as the
decoupling hypersurface is approached. This behavior is shown in Figure 3
depicting the ratios $\tau/t_{H}$ and $t_{\gamma}/t_{H}$
as well their dependence on $|\Delta_{i}^{(s)}|$.

\subsection{The truncated transport equation}

In many cases, just for the sake of mathematical simplicity, the transport
equation (\ref{pitensor}) governing the  evolution of the dissipative
stress tensor has been replaced in by the more manageable
Maxwell--Cattaneo--like expression, namely the so called ``truncated''
transport equation
\begin{equation}
\tau \dot\Pi_{cd}\,h^c_ah^d_b+\Pi_{ab} +2\eta\,\sigma_{ab}= 0.
\label{namely}
\end{equation}
\noindent
an equation still complying with relativistic causality and
stability in the sense mentioned earlier, and justified
when the four--dimensional gradient
of the combination $\tau u^{c}/(T \eta)$ becomes negligibly small.
The models presented here provide a theoretical context in which to
test how suitable can this truncated transport equation be for studying the
evolution of an interactive matter--radiation mixture. To do this consider
the relaxation time that follows from (\ref{namely})
\begin{equation}
\tau_{mc} = - \frac{P}{\dot{P} + \frac{2}{5} p \sigma}
= \frac{4 + 3 \Delta_{i}^{(r)} - 4 \Gamma}{4 \frac{\dot{y}}{y}
[4 + 3 \Delta_{i}^{(r)} - 4 \Gamma] - \frac{1}{5} \sigma[31(4+
3 \Delta_{i}^{(r)}) - 15 \Gamma]},
\label{taumc}
\end{equation}
\noindent
where $\tau_{mc}$ is the relaxation time associated with the
Maxwell--Cattaneo equation. Comparing (\ref{taumc}) with the relaxation time
(\ref{tau}) that emerges from the full theory (i.e. from (\ref{pitensor}))
\begin{equation}
\tau = - \frac{P}{\dot{P} + \frac{2}{5}p \sigma - \frac{\dot p}{p} P}
= - \frac{1}{4 \sigma} \frac{(4 + 3 \Delta_{i}^{(r)} -4 \Gamma)
(4 + 3 \Delta_{i}^{(r)} - \Gamma)} {\frac{4}{5}
[4 + 3 \Delta_{i}^{(r)} - 4 \Gamma]^{2}+ \frac{171}{256} \Gamma^{2}}.
\label{tauf}
\end{equation}
\noindent
we can see that $\tau = \tau_{mc} (1 + \tau_{mc} (\dot{p}/p))^{-1}$,
therefore since $\dot{p}/p < 0$ (as long as $\Psi > 0$, which is always
the case) we have $\tau > \tau_{mc}$. Now two possibilities arise:
\begin{enumerate}
\item
\underline{Adiabatic condition} $\Delta_{i}^{(s)} = 0$, in this
case Figure 4 shows that $\tau_{mc} < t_{H}$ for the whole evolution.

\item
\underline{Quasi--adiabatic condition}  $\Delta_{i}^{(s)} \neq 0$, in
this case as seen from Figure 5, $\tau_{mc}$ does not diverge nor overtakes
$t_H$ at any time in the whole evolution, while $\tau$ diverges and
overtakes $t_H$ for $y > y_{D} \approx 10^{2.4}$.
\end{enumerate}

\noindent
This together with the fact that close to the decoupling surface $t_{\gamma}$ 
becomes larger than $\tau_{mc}$  it follows that the Maxwell--Cattaneo 
equation (\ref{namely}) applied to the LTB models derived here and 
under the constraints imposed  on the initial density contrasts 
($\Delta_{i}^{(r)} \approx 10^{-4}$ and 
$10^{-9} < |\Delta_{i}^{(s)}| < 10^{-7}$) by the observed CMB 
anisotropies, does not provide a relaxation time that behaves
reasonably in conditions near the matter--radiation  decoupling.

It is widely  assumed nowadays that the bulk of the matter of the Universe 
(about ninety percent) is not baryonic but belongs to some hitherto unknown
form of ``dark matter" (see e.g. \cite{bahcall}) -though voices of dissent 
can be heard \cite{rowan}. We have considered a baryon--photon mixture and 
so have not taken this into account in our calculations. However, we can
roughly verify to what extent our results would change had we incorporated
some form of this missing matter. If we assume that it is mostly 
non--relativistic and that it amounts to $20$ times the baryon abundance, 
the inclusion of such missing matter corresponds to decreasing the value 
of the parameter $\epsilon$ by a factor of $20$ (from $10^{3}$ down
to $50$). As shown in Figure 6, carrying on this modification does not 
imply a dramatical change in our results. Further, it is worthwhile 
remarking that the introduction of a small cosmological constant
or a quintessence fluid would not make sense for the models we are
dealing with. Since these forms of ``dark energy" contributions (as 
it is nowadays in vogue) are used to explain the alleged accelerated 
expansion of the present day Universe, they should be sub--dominants
at the epcoch contemplated in this paper (see e.g. \cite{quintessence}).

\section{Concluding remarks}

It is worthwhile mentioning that the main results of this paper: (a)
inadequacy of the truncated equation (\ref{namely}) to provide a
relaxation time that behaves reasonably near matter-radiation
decoupling, and (b) that $\tau$ is much larger than 
$t_{\gamma}$ for most of the  radiative era, strictly apply for the 
specific model examined in \cite{mixt}, based on the LTB metric. 
These results also rely on the expression of $\tau$ obtained in terms 
of the transport equations (\ref{pitensor}) and (\ref{namely}). However, 
we argue that in spite of these limitations these results can complement 
and enrich the existing discussion in the literature concerning the 
adequacy of truncated transport equations of dissipative stresses in 
a cosmological context -see \cite{discussion} and references therein.
Our analysis suggests that, in any case, these equations should be 
used with  due care.  We believe it is important to stress that 
(as far as we are aware) these important features of the equations 
of irreversible thermodynamics have not been examined previously in 
the context of inhomogeneous metrics able to accommodate dissipative 
shear stresses.

\acknowledgements
We thank the anonymous referees for constructive criticisms leading to 
an upgrading of the paper. This work has been partially supported 
by the Spanish Ministry of Science and Technology and the National 
University of Mexico (UNAM), under grants BFM 2000-0351-C03-01 and 
DGAPA-IN-122498, respectively.

\  \\
\  \\
\newpage
\section*{List of caption for figures}
\noindent
{\bf Figure 1}\\
The plot depicts the functions $\Psi$ and $\Gamma$, given by (6) and (7), in
terms of $\log_{10}(y)$ and $\log_{10}(|\Delta_{i}^{(s)}|)$, from the
initial hypersurface $y = 1$ corresponding to $T_{i} = 10^{6}$ Kelvin. Notice
how these functions are almost unity for all the evolution range.\\

\noindent
{\bf Figure 2}\\
Logarithmic plot of $\tau$ {\it vs} $|\Delta_{i}^{(s)}|$ for $y=y^{2.5}$
corresponding to $T=T_{_D}$. Note that $\tau$ diverges for $|\Delta_{i}^{(s)}|
\approx 10^{-8}$.  Since $t_{\gamma}$ also diverges as $T \to T_{_D}$, the 
behavior of $\tau$ for this value of $|\Delta_{i}^{(s)}|$ is reasonable.\\

\noindent
{\bf Figure 3}\\
Logarithmic plot depicting the ratios $\tau/t_{_H}$ (shaded surface) and
$t_{\gamma}/t_{_H}$ (non-shaded surface) {\it vs} the parameters 
$log_{10}(y)$ and $|\Delta_{i}^{(s)}|$. The former ratio is far larger 
than the latter for most of the evolution range. For
$|Delta_{i}^{(s)}| \approx 10^{-8}$ the two ratios become comparable
in magnitude near the decoupling surface ($t_{\gamma}/t_{_H} = 1$).

\noindent
{\bf Figure 4}\\
Logarithmic plots of the ratios $\tau/t_{H}$ (top surface) and
$\tau_{mc}/t_{_H}$ (bottom surface) {\it vs} the parameters $y$ and
$|\Delta_{i}^{(r)}|=(4/3)|\Delta_{i}^{(m)}|$ under adiabatic conditions
($|\Delta_{i}^{(s)}| = 0$). Neither one of these relaxation times diverges
nor overtakes $t_{H}$, therefore they are not physically realistic near
the decoupling surface $t_{\gamma}/t_{H} = 1$.
\\

\noindent
{\bf Figure 5}\\
The same plot as figure 4 but under quasi--adiabatic
conditions. The ratio $\tau/t_{H}$ diverges for $ log_{10} y \approx 2.4$
while the the ratio $\tau_{mc}/t_{H}$ remains finite for all $y$, therefore
the full transport equation yields a reasonable relaxation time while
the truncated equation does not.\\

\noindent
{\bf Figure 6}\\
This figure compares $\tau/t_{_H}$ (A), $\tau_{mc}/t_{_H}$ (B) and
$t_{\gamma}/t_{_H}$ (C) {\it vs} $\log_{10}(y)$. The thick curves were 
obtained assuming $T=10^6$ Kelvin and an initial photon to baryon ratio
$n_i^{(r)}/n_i^{(m)}=10^9$, hence $\epsilon \approx \rho_i^{(r)}/\rho_i^{(m)}
= 10^3$. If we assume that CDM does not interact with the photon gas and that
it constitutes most of non--relativistic matter, say 20 times the abundance of
baryons, then we would need to use $\epsilon=50$, lower by a factor of $20$.
The thin curves were obtained using this value of $\epsilon$ with
$|\Delta_{i}^{(s)}|=10^{-8}$ (upper curves) and $|\Delta_{i}^{(s)}|=10^{-9}$
(lower curves). Notice that these two curves are indistinguishable in (C),
thus indicating that $t_{\gamma}$ is rather insensitive to changes in
$|\Delta_{i}^{(s)}|$.
The changes in $\epsilon$ due to the inclusion of the rest--mass of CDM 
are also minimal, shifting the best value of $|\Delta_{i}^{(s)}|$ to 
an order of magnitude less.

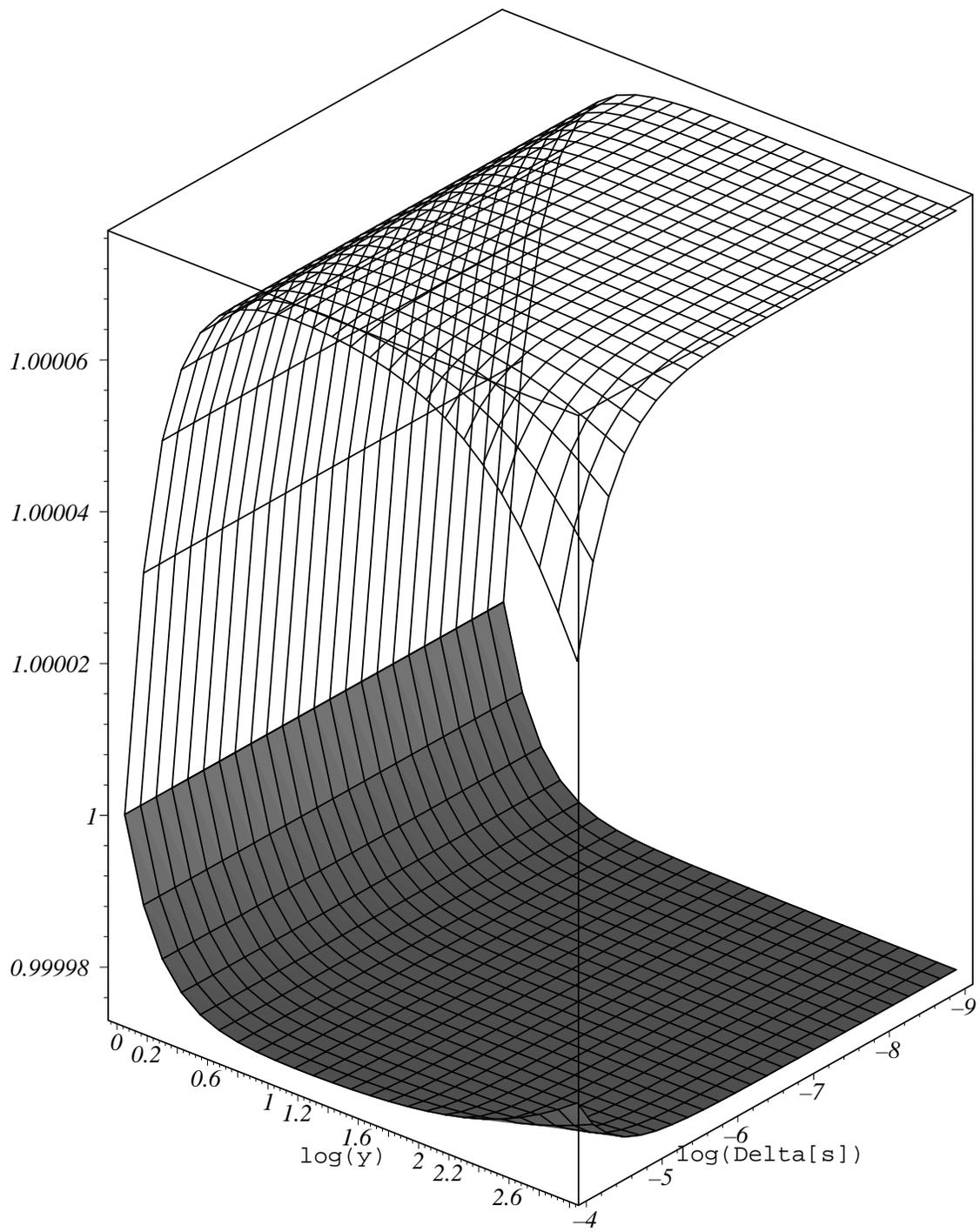
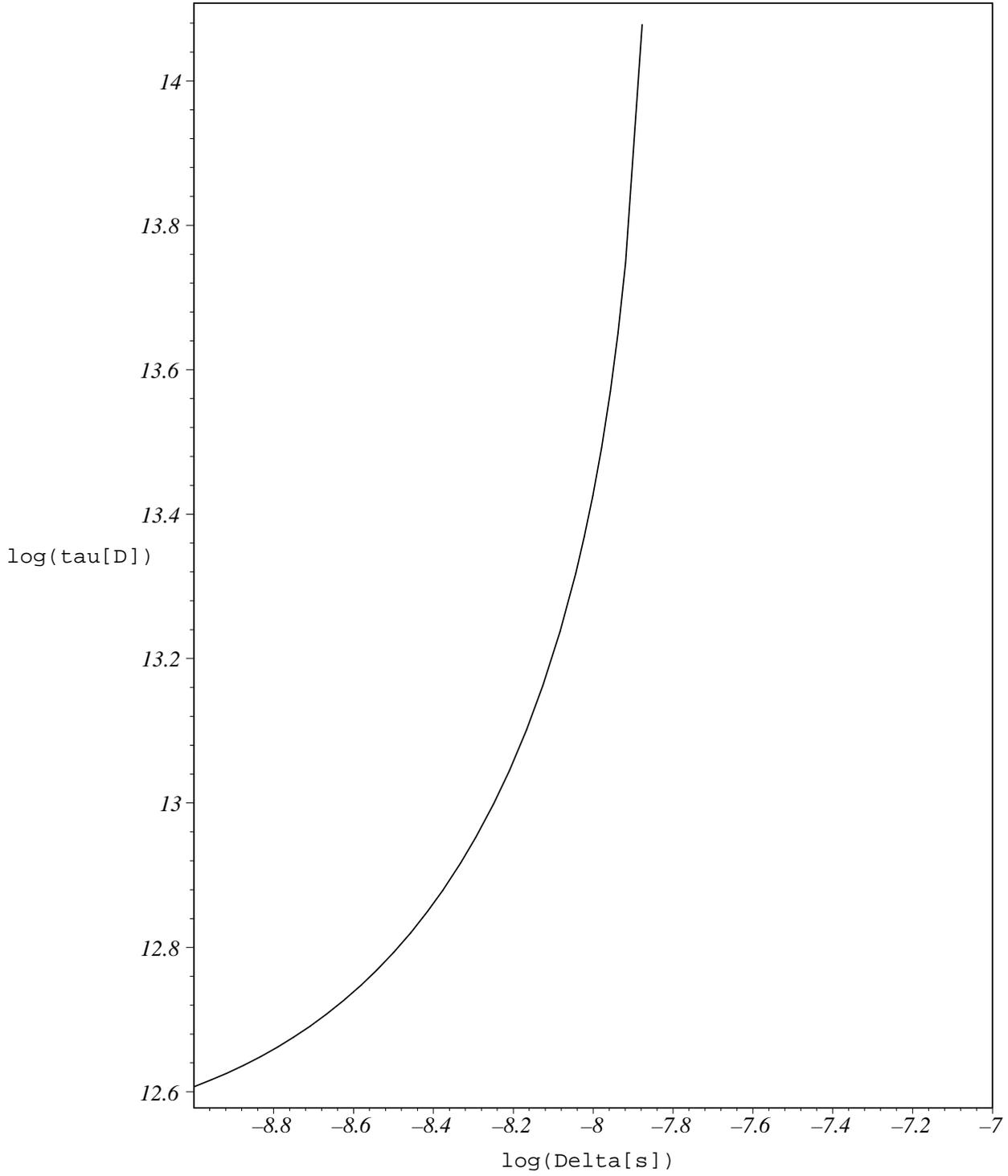
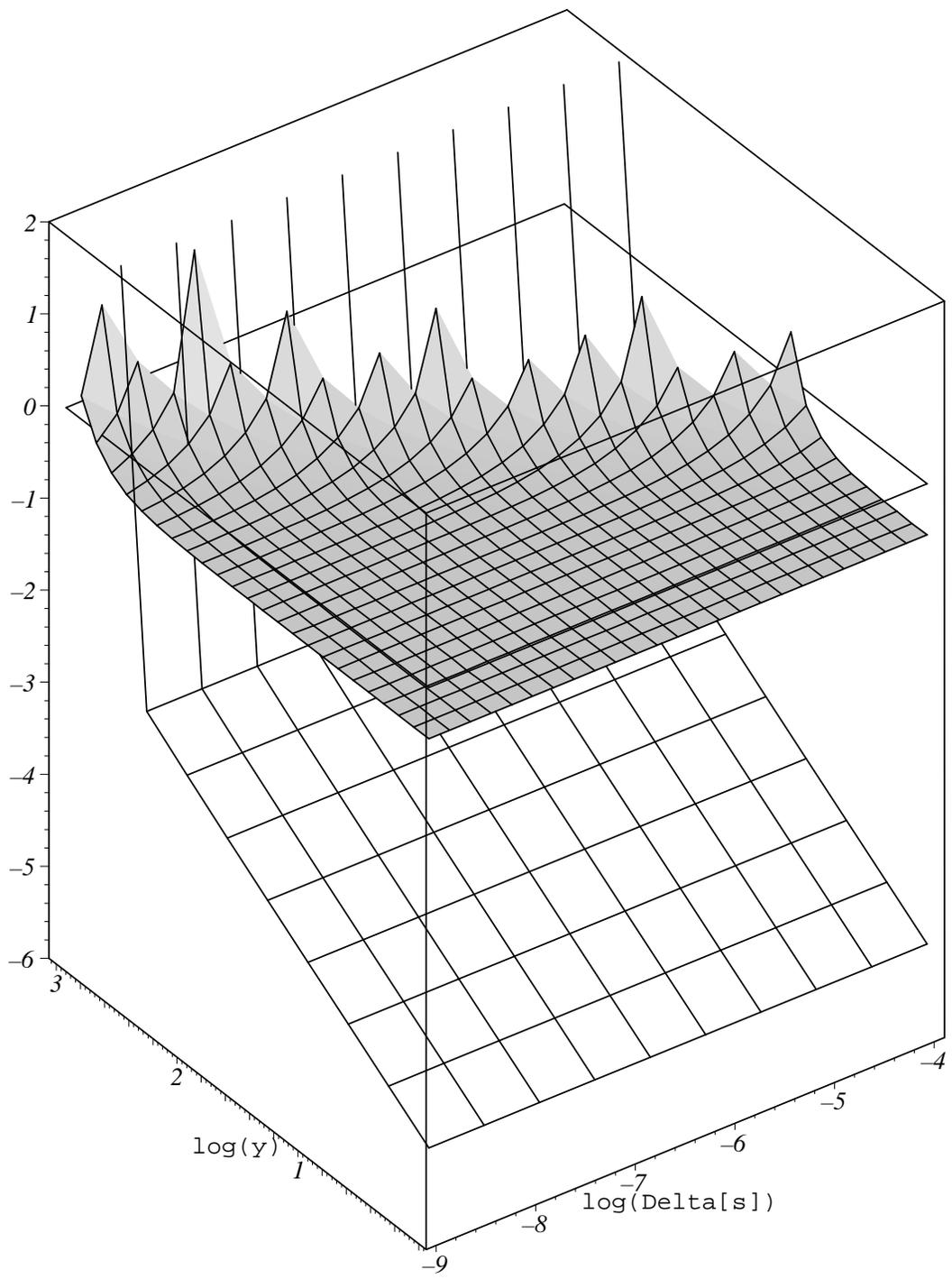
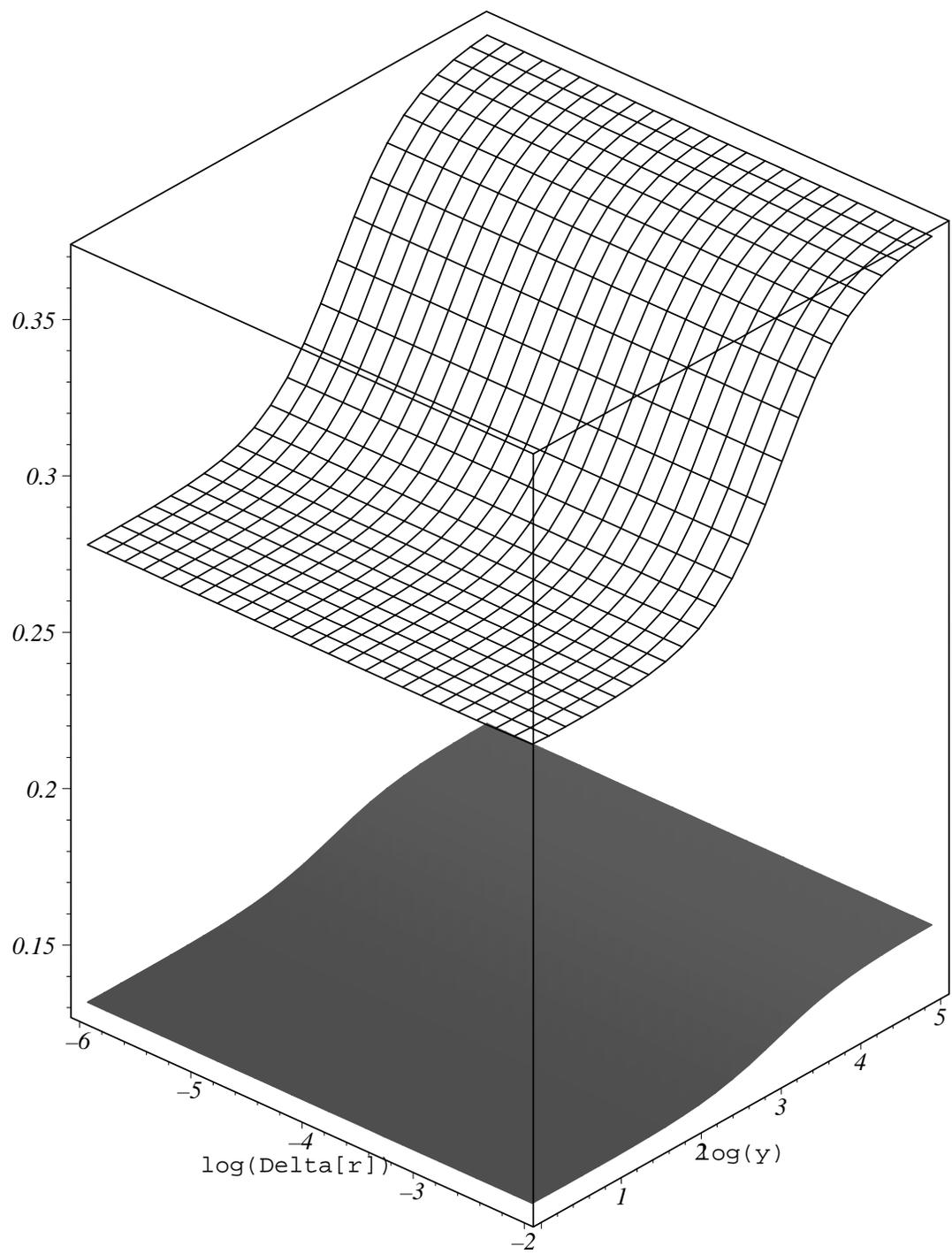
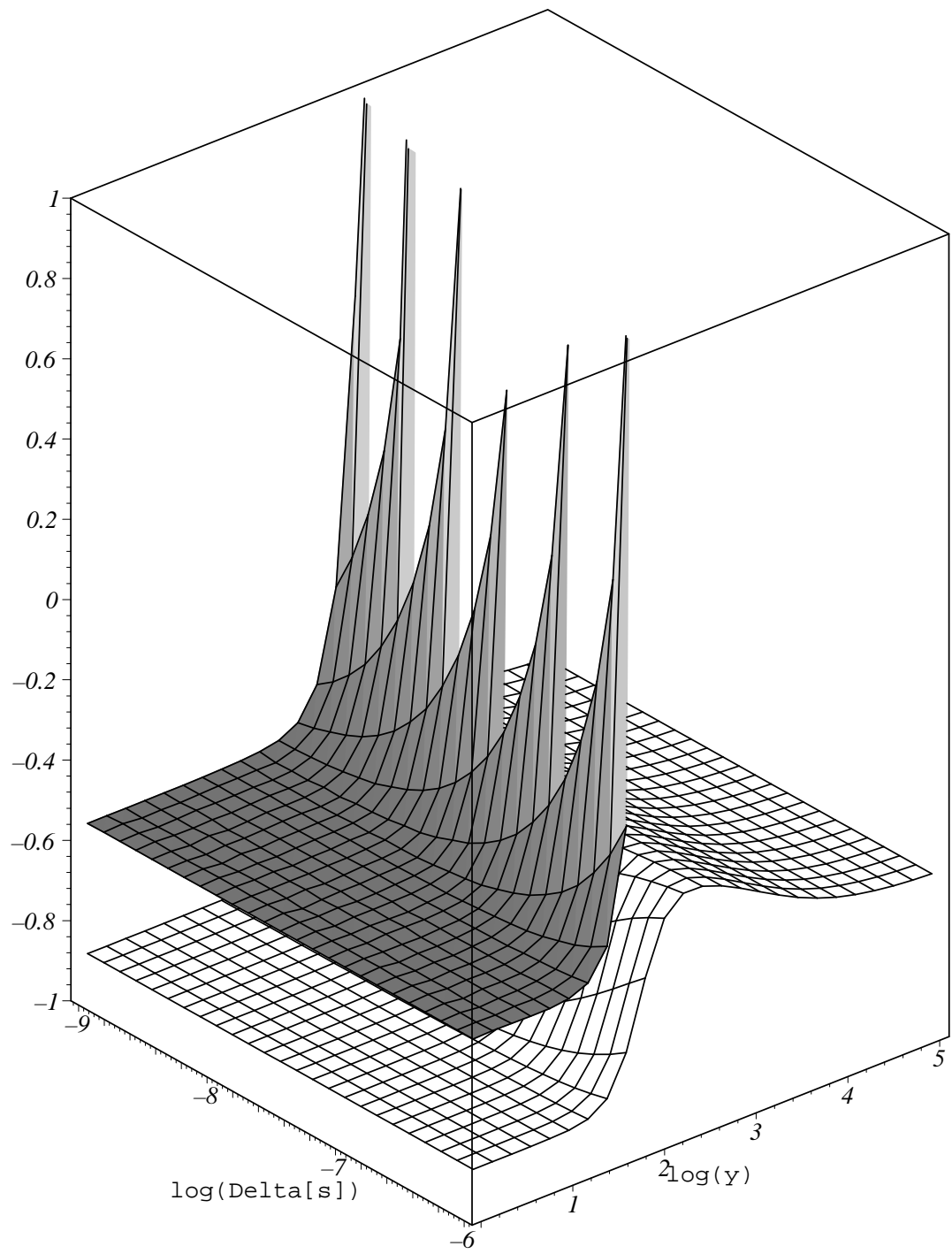
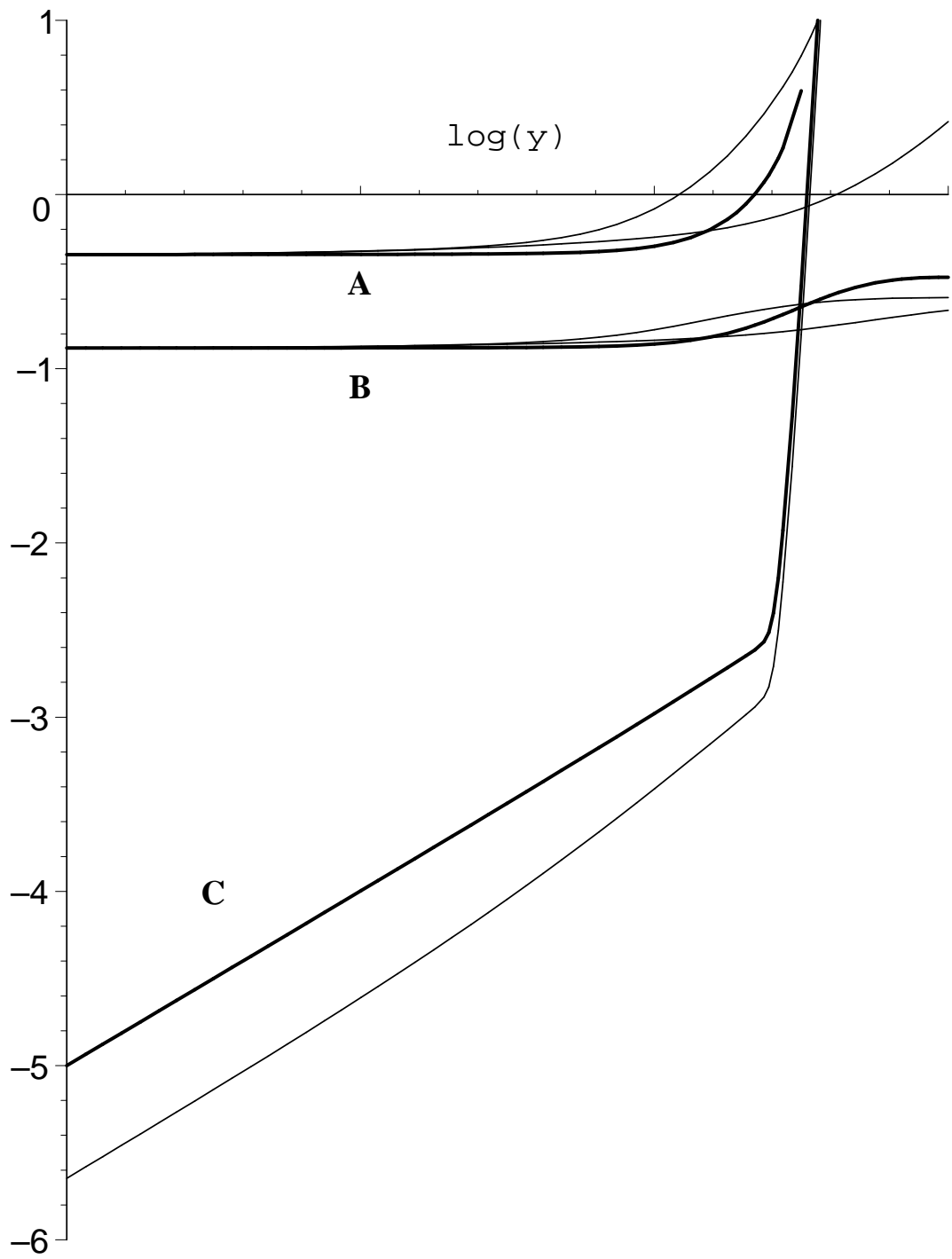
\end{document}